\documentclass[proceedings]{JHEP}
\usepackage{amssymb}
\usepackage{latexsym}
\usepackage{verbatim}
\usepackage{epsfig}
\conference{Corfu Summer Institute on Elementary Particle
Physics, 1998}
\title{Quantum and Braided Integrals}
\author{Chryssomalis Chryssomalakos\footnote{%
Present address:
Instituto de Ciencias Nucleares, Universidad Nacional Aut\'onoma
de Mexico, Apdo. Postal 70-543, 04510 M\'exico, D.F., M\'exico,
\texttt{chryss@nuclecu.unam.mx}%
}\\
Department of Theoretical Physics\\
University of Valencia\\
and\\
I.F.I.C.\\
University of Valencia--CSIC\\
E--46100 Burjassot (Valencia)\\
Spain\\
\email{chryss@lie3.ific.uv.es}
}
\abstract{%
We give a pedagogical introduction to integration techniques
appropriate for non-commutative spaces while presenting some new
results as well. A rather detailed
discussion outlines the motivation for adopting the Hopf algebra
language. We then present some trace formulas for the integral on
Hopf algebras and show how to treat the $\int 1=0$ case. 
We extend the discussion to braided Hopf algebras
relying on 
diagrammatic techniques. The use of the general formulas is
illustrated by explicitly worked out examples.
}
\newtheorem{exatitle}{Example}[section]
\newenvironment{example}[1]%
{\begin{exatitle}#1 \end{exatitle}}%
{\hfill $\Box$ \\}
\newcommand{\be}{\begin{equation}}
\newcommand{\ee}{\end{equation}}
\newcommand{\bae}{\begin{eqnarray}}
\newcommand{\eae}{\end{eqnarray}}
\def\bb{\bibitem}
\def\nn{\nonumber}
\def\ff{\nn \\}
\def\A{{\mathcal A}}

\def\CC{{\mathbb C}}

\def\E{{\mathcal E}}
\def\bE{\bar{\E}}

\def\H{{\mathcal H}}

\def\R{{\mathcal R}}

\def\T{{\mathcal T}}
\def\U{{\mathcal H}}

\def\id{\mathop{\rm id}}
\def\ot{\otimes}

\def\tr{\triangleright}
\def\ra{\rightarrow}

\def\ip#1#2{\left\langle #1, #2\right\rangle}
\def\inprod#1#2{\left\langle #1, #2\right\rangle}

\def\vaca{\Omega_{\A}}
\def\lva{\langle \Omega_{\A} |}
\def\rva{| \Omega_{\A} \rangle}
\def\vacu{\Omega_{\U}}
\def\olo#1{\langle #1 \rangle}
\def\Lolo#1{\langle #1 \Lar}
\def\Rolo#1{\langle #1 \Rar}
\def\scr{\scriptscriptstyle}
\def\Rar{\rangle^{\scr R}}
\def\Lar{\ar^{\scr L}}
\def\trRar{\Rar_{tr}}
\def\uA{1_{\A}}
\def\uU{1_{\H}}
\def\ar{\rangle}
\def\urv{| \Omega_{\U} \rangle}
\def\ulv{\langle \Omega_{\U} |}
\def\lva{\langle \Omega_{\A} |}
\def\rva{| \Omega_{\A} \rangle}
\def\fe{& = &}
\def\la{\langle}
\def\Lar{\ar^{\scr L}}
\def\Rolo#1{\langle #1 \Rar}
\def\Rda{\delta^{\scr R}_{ \mathcal A}}

\def\hR{\hat{R}}
\def\xs{\xi_{i_{1 \dots k}}}
\def\sx{\sigma_{i_{k\dots 1}}}
\def\eg{\hbox{\it e.g.}}
\def\etc{\hbox{\it etc.}}
\def\ie{\hbox{\it i.e.}}
\begin{document}
\section{Introduction}
\label{Intro}
It is a recurrent theme of recent years that the physics of
the not-so-distant future will probably involve some species
of `quantum space', a fuzzy substratum free of the singularities
inherent in the classical concept of a point. Then, if a long tradition
continues, the algebraic codification of its properties 
will entail non-commutativity as an essential ingredient.  
The first textbooks on
this new physics will probably include an appendix with 
elements of integration techniques on the relevant non-commutative
spaces.  We know little yet about what these spaces might turn
out to be but there
are general arguments that the coordinate functions over them
generate Hopf algebras. While
contemplating on the rest of the  contents of the book, we attempt 
here a first sketch of a part of its appendix---a short, informal
crash course in some techniques in quantum integration. 

Sec. \ref{FDHA} motivates the algebraic setting chosen. Hopf algebras
are close enough to classical groups to guarantee  a continuity
of language and yet accommodate
naturally rather exotic geometrics. Assuming a
convinced reader, Sec. \ref{QInt} then goes on with the basics
of Hopf algebra integration, while Sec. \ref{HASBHA} and
\ref{BInt} extend 
the discussion to the braided case. We rely on
simple, detailed examples to illustrate the proposed techniques.
\section{Quantum Points + Translations\\= Hopf Algebras} 
\label{FDHA}
\subsection{Quantum points} 
\label{Qp}
Once we have
decided to abandon classical manifolds we have to make up our
mind on what kind of spaces we would like to consider. It is
common place by now to emphasize that, having admitted 
non-commutativity in the algebra of
functions over the `space' under consideration, we can no longer
talk about an underlying manifold consisting of points---what we are 
left with is the algebra of
functions itself.  What needs perhaps to be also stressed, to
dispel a certain feel of unease that comes with a pointless space, 
is that, as we will see in the case of Hopf algebras, 
the `manifold' 
still consists of well-defined entities, which one could think
of as `quantum points'. To make this more precise, consider a
non-commutative, in general, algebra 
$(\A, \, m, \, \uA)$, where $\A$ is a vector space and $m$ a  
multiplication map, 
which
will play in the sequel the role of the algebra of functions over
some `quantum manifold'---we will denote its elements by $a$, $b$,
\etc; $\uA$ is the unit function.  We concetrate on the conceptual
aspects of the problem and take $\A$ to be
finite-dimensional, avoiding potential divergences of the sums
that enter in our discussion. Nevertheless, we find the
picture of a continuous group manifold invaluable in developing
some intuition and, although the latter involves an
infinite-dimensional algebra of functions, we will keep it in the
back of our mind as a guide. 
Dual to $\A$, in a sense to be made precise shortly,
is a `quantum space' $\H$, 
with elements $g$, $h$, $x$, $y$ \etc---these 
are the `quantum points' referred to
earlier and among them there is an identity $\uU$. Functions 
evaluated on points give numbers, {\em even in the
quantum case}. We formalize this by introducing an {\em inner
product}, or {\em pairing},  
$\ip{\cdot}{\cdot}$ between $\A$ and $\H$, with values 
in $\CC$, via
$\ip{h}{a} \equiv a(h)\in \CC$ for any function $a \in \A$ and `point'
$h \in \H$.

One might ask at this stage, how can functions
valued in $\CC$ be non-commutative? The answer lies actually in
the dual. To see this, consider a classical point $h_{cl}$ and
evaluate on it the product of functions $ab$:
$(ab)(h_{cl})=a(h_{cl})b(h_{cl})$ which is equal to
$(ba)(h_{cl})$. What happens is that when a classical point sees a 
product of functions, it `splits' in two copies of itself, $h_{cl}
\rightarrow h_{cl} \otimes h_{cl}$, and feeds each
of them as argument to each of the factors in the product
\bae
(ab)(h_{cl}) & \equiv &
(a \otimes b)( h_{cl} \otimes h_{cl})
\ff
 & = & a(h_{cl})b(h_{cl})
\, . 
\label{copcl}
\eae
If the underlying manifold consists in its
entirety of such classical points, the functions $ab$ and $ba$
agree when evaluated on all points and can (and should) therefore be
considered equal. The conclusion is  that the function algebra over a
classical space is commutative because the classical {\em
coproduct} map 
\be
\Delta: \H \ra \H \otimes \H; 
\qquad 
h_{cl} \mapsto h_{cl} \otimes h_{cl} 
\label{Dcl}
\ee
is symmetric under the exchange of its two tensor factors. 
To put our notation in some use,
we rewrite~(\ref{copcl})
\bae
\ip{h_{cl}}{ab} & \equiv & \ip{h_{cl}}{m(a \ot b)} 
\ff
 &=& \ip{\Delta(h_{cl})}{a \otimes b}
\ff
 &=& \ip{h_{cl} \otimes h_{cl}}{a \otimes b} 
\ff
 &=& \ip{h_{cl}}{a} \ip{h_{cl}}{b}
\ff
 &=& a(h_{cl})b(h_{cl})
\, .
\label{ipdet}
\eae
Notice how, in the second line above, the coproduct map
$\Delta$ in $\H$ is dual to the product map $m$ in $\A$. 
We see that, in
some sense, classical points are quite primitive, in that the
only information they carry is about their own position---when
confronted with products of functions they can only produce
multiple copies of themselves. 
Quantum points can do better than this. When paired with products
of functions they split, via a coproduct map as above, 
in two other quantum points, $h \mapsto \Delta(h) \equiv 
h_{(1)} \ot h_{(2)}$, with $h_{(1)} \ot h_{(2)} \neq h_{(2)} \ot
h_{(1)}$ in general, and feed each of them in the two factors of
the product
\bae
\ip{h}{ab} &=& \ip{h_{(1)}}{a} \ip{h_{(2)}}{b}
\ff
 & \neq & \ip{h_{(2)}}{a} \ip{h_{(1)}}{b}
\ff
 &=& \ip{h}{ba}
\, .
\label{habq}
\eae
Consequently, $ab \neq ba$ in general. By examining all the
values of all the functions in $\A$, one can nevertheless establish 
commutation relations between them, \eg \  if $ba$ systematically
returns twice the value of $ab$, on all quantum points, one
imposes the relation $ba = 2ab$ in the algebra. The rule for
assigning the two points $h_{(1)} \ot h_{(2)}$ to $h$ (\ie \ the
coproduct $\Delta(h)$) cannot of
course be arbitrary. If the product in $\A$ is to be associative, 
$\Delta$ has to be {\em coassosiative}
\be
(\Delta \ot \id) \circ \Delta = (\id \ot \Delta) \circ \Delta
\, .
\label{coasso}
\ee
The identity point is taken to be classical (or {\em grouplike}):
$\Delta(\uU) = \uU \ot \uU$. 
We  introduce also a {\em counit} $\epsilon$ in $\H$. $\epsilon(h)$ is
defined to be the value of the unit function on $h$:
$\epsilon(h)\equiv \ip{h}{\uA}$. This can be different from 1 since
$h$ can be an arbitrary linear combination of elements in $\H$. 
Just like $(\A, \, m, \, \uA)$ defines an algebra, the triple $(\U,
\, \Delta, \, \epsilon)$ defines a {\em coalgebra}. Notice that
we don't have, at this point, any notion of product of quantum
points---$\H$ is not yet an algebra (and, similarly, $\A$ is not
yet a coalgebra). 

Experimenting a little with the above, one discovers that
$\Delta(h)$ has, in general, to involve a sum over pairs of points,
rather than a single pair. We continue to denote such a sum by
$h_{(1)} \ot h_{(2)}$ \ie
\be
\Delta(h) = \sum_{i} h_{(1)}^i \ot h_{(2)}^i \equiv h_{(1)} \ot
h_{(2)}
\, .
\label{Swee1}
\ee
The coassociativity mentioned above says that
\bae
h_{(1)(1)} \ot h_{(1)(2)} \ot h_{(2)}
&=& h_{(1)} \ot h_{(2)(1)} \ot h_{(2)(2)}
\ff
 &\equiv & h_{(1)} \ot h_{(2)} \ot h_{(3)}
\, ,
\ff
 & &  
\label{coasso2}
\eae
where in the last line we have renumbered sequentially the
subscripts, given that the particular order of applying the
successive $\Delta$'s does not matter. There are two points that
should be kept in mind to avoid confusion with the above notation: 
first, $h_{(1)} \ot h_{(2)}$ in~(\ref{Swee1}) and in the second
line of~(\ref{coasso2}) 
denotes two different things and second, there are $n-1$ implied
summations in $h_{(1)} \ot \dots \ot h_{(n)}$. 
\subsection{Translations} 
\label{Tran}
We continue building up an arena for physics by introducing
a concept of translations for our functions---it is in terms of
this that the invariance of our integral will be expressed (there
are of course other uses for it as well).
Quite generally, one can describe (left) translated functions by
introducing a new algebra $\T$ that acts on the points in $\H$
via a {\em left action} 
\bae
\tr: \quad \T \ot \H & \ra & \H 
\, ,
\ff
 t \ot h & \mapsto & t \tr h 
\, ,
\label{lact}
\eae
\ with 
\be
(tu) \tr h = t \tr (u \tr h)
\, ,
\label{actdef}
\ee
 $t,u \in
\T$ and $h \in \H$. Then the translated function $a \in \A$  by
$t$, which we write as $a_t$, is defined via $a_t(h) 
\equiv a(t \tr h)$. 
The simplest choice for $\T$ is $\H$ itself with $\tr$ a
left multiplication, in other words we can endow $\H$ with an
associative 
product (turning it into an algebra) and define $a_h(g) \equiv
a(hg)$. From this we can abstract the notion of an `indefinitely
translated' function $a_{(\cdot)}(\cdot)$, where the argument in
the subscript defines the translation and the second argument
evaluates the translated function. Such a function can be written
as a sum over tensor products of single-argument functions
\bae
a_{(\cdot)}(\cdot) &\equiv & 
\sum_i a_{(1)}^i(\cdot) \ot a_{(2)}^i(\cdot)
\ff
 & \equiv &
a_{(1)} \ot a_{(2)}
\, ,
\label{copfun}
\eae
which we realize as a coproduct in $\A$.
Coassociativity of this coproduct is dual to the associativity
of the product in $\H$, which itself guarrantees the general
property~(\ref{actdef}) of an action. $\Delta$ can equally well
be thought of as describing translations from the right, in this
case the element of $\H$ that describes the translation is fed in
the second tensor factor and the resulting right-translated
function accepts arguments in the first. 
We can also introduce a counit $\epsilon$ in $\A$: $\epsilon(a)$ is 
the value of $a$ at the identity. 
Since the identity point is classical, $\epsilon$ supplies
a one-dimensional representation for $\A$: $\epsilon(ab)
=\epsilon(a) \epsilon(b)$. 
Notice that,  again, product and coproduct are dual
\be
\ip{hg}{a} = \ip{h \ot g}{a_{(1)} \ot a_{(2)}}
\, .
\label{pcpd}
\ee
It is rather natural to impose on translations a certain
covariance property: they should respect the `quantum nature' of
$\A$ \ie \ the `indefinitely' translated functions should obey the same
commutation relations as the original ones. This implies that the
coproduct should be an algebra homomorphism
\bae
\Delta(ab) &=& \Delta(a)\Delta(b) 
\ff
(ab)_{(1)} \ot (ab)_{(2)} &=& a_{(1)}b_{(1)} \ot a_{(2)} b_{(2)}
\, .
\ff
\label{cophom}
\eae
By looking at $\ip{hg}{ab}$ one finds that the same should hold
in $\H$---this turns $\A$ and $\H$ into {\em bialgebras}. 
What is missing in order to turn both of them  into
full-fledged {\em Hopf algebras} is the {\em antipode} $S$. This
can be useful if, after translating our functions by some
$h$, we change our mind and wish to undo the
translation. When
$\H$ is a classical group (discrete or not), $S(h)=h^{-1}$ and
$\A$ inherits an antipode via $S(a)(h) \equiv a(S(h))$. In the
general case we define $S$ via 
\be
S(h_{(1)}) h_{(2)} = \epsilon(h) = h_{(1)} S(h_{(2)})
\label{Sdef}
\ee
(similarly for $\A$) and it still holds 
\be
\ip{S(h)}{a} = \ip{h}{S(a)}
\, .
\label{Sdual}
\ee
$S$ is an antihomomorphism, $S(hg)=S(g)S(h)$. 
Notice that $S^2$ is not necessarily the identity map; this 
can be a nuissance in transcribing classical results that hold
for groups in the general Hopf algebra case but it is also the source
of unexpected novelties. We will often
pick a {\em linear} basis
$\{f^i\}$, $i=0,\dots,N$ in $\A$. This means that any function
in $\A$ can be written as a linear combination, with coefficients in
$\CC$, of the $f^i$---we choose $f^0 = \uA$, the unit function.
Similarly, $\{e_i\}$, $i=0,\dots,N$, will be a dual linear 
basis in $\H$ with $\ip{e_i}{f^j}={\delta_i}^j$; $e_0$ will
denote the identity $\uU$. 

To summarize, we started from a primitive notion of `quantum
space' and the functions defined over it and saw that, in
general, the former forms a coalgebra while the latter an
algebra. Further introducing translations and requiring their
compatibility with the algebra of functions results in a 
symmetric structure, turning both the space and
the functions over it into bialgebras. Adding a notion of inverse
we end up with a pair of dual Hopf algebras.
We illustrate the
above concepts in the following two examples. 
\begin{example}{Discrete classical space}
\label{dcs}
$\H$ in this case is a discrete group algebra. It contains $n$ 
classical points 
$\{e_i\}$, $i=0, \dots,n-1$ and their linear combinations.
The coproduct is $\Delta(e_i)=e_i \ot e_i$ (no
summation over $i$). $(\Delta \ot \id) \circ \Delta(e_i)$ is
given by $e_i \ot e_i \ot e_i$ (no summation) and agrees with 
$(\id \ot \Delta) \circ \Delta(e_i)$. All
products $e_ie_j$ are contained in the set $\{e_i\}$, hence $\H$
is finite-dimensional, linearly spanned
by $\{e_i\}$. The group law can then be given in terms of the
constants ${M_{ij}}^k$ via $e_i e_j={M_{ij}}^k e_k$. 
The role of derivatives is played by the difference operators
$e_{ij} \equiv e_i -e_j$, $i<j$, with $\Delta(e_{ij})=e_{ij} \ot
1 + 1 \ot e_{ij}$ and $\epsilon(e_{ij})=0$, $S(e_{ij})=-e_{ij}$.  

$\A$
is generated  by the functions $\{f^i\}$, $i=0, \dots,n-1$,  with
$\ip{e_i}{f^j}={\delta_i}^j$, and is commutative. $f^i$ is
is a delta-like function peaked over $e_i$, hence $f^i f^j=0$
whenever $i \neq j$ while each $f^i$ squares to itself. Since no
new functions can be produced by multiplication, $\A$ is linearly
spanned by the set $\{f^i\}$. The coproduct in $\A$ is given by
the same numerical constants that give the product in $\H$:
$\Delta(f^i)={M_{jk}}^i f^j f^k$.   
Notice that $(e_i \ot f^i)(e_j \ot f^j) = e_i \ot e_i \ot f^i$,
which in turn is equal to $\Delta(e_i) \ot f^i$. $e_i \ot f^i$ is
the {\em canonical element} and the above identity is in the
spirit of $e^{a+b}=e^ae^b$. 

The unit function is 1 on
every $e_i$ so that $\epsilon(e_i)=1$. All $f^i$, except $f^0$,
vanish on the identity hence $\epsilon(f^0)=1$,
$\epsilon(f^i)=0$, $i \neq 0$. For the antipode we have
$S(e_i)={e_i}^{-1}\equiv {S_i}^je_j$ and $S(f^i)={S_j}^i f^j$. 
\end{example}
\begin{example}{A discrete `quantum space'}
\label{dqs}
$\H$ is generated by $1$, $x$ and $y$ with $x^2=0$, $y^2=y$ and
$yx=-xy+x$. The coproduct  is 
\bae
\Delta(x) &=& x \ot 1 + 1 \ot x -2y \ot x 
\ff
\Delta(y) &=& y \ot 1 + 1 \ot y -2 y \ot y
\, ,
\label{Hopfdqs}
\eae
\ie \, $x$, $y$ are of the difference operator type. The rest of the
Hopf structure is $\epsilon(x)=\epsilon(y)=0$ and
$S(x)=-(1-2y)x$, $S(y)=y$ (notice that $S^2(x)=-x$). $\H$ is spanned 
linearly by
$\{e_i\}=\{1,x,y,xy\}$. Writing $e_i e_j={(M_i)_j}^k e_k$ and
$\Delta(e_i)={(W^k)_i}^j e_j e_k$ we find
\be
\begin{array}{rclcrcl}
M_0 & \! = \! & \left(
\begin{array}{cccc}
1 & 0 & 0 & 0 \\
0 & 1 & 0 & 0 \\
0 & 0 & 1 & 0 \\
0 & 0 & 0 & 1
\end{array}
\right)
& & M_1 & \! = \! & \left(
\begin{array}{cccc}
0 & 1 & 0 & 0 \\
0 & 0 & 0 & 0 \\
0 & 0 & 0 & 1 \\
0 & 0 & 0 & 0 
\end{array}
\right)
\\
& & & & & & 
\\
M_2 & \! = \! & \left(
\begin{array}{cccc}
0 & 0 & 1 & 0 \\
0 & 1 & 0 & -1 \\
0 & 0 & 1 & 0 \\
0 & 0 & 0 &0 
\end{array}
\right)
& & M_3 & \! = \! & \left(
\begin{array}{cccc}
0 & 0 & 0 & 1 \\
0 & 0 & 0 & 0 \\
0 & 0 & 0 & 1 \\
0 & 0 & 0 & 0 
\end{array}
\right)
\end{array}
\label{MWdqs}
\ee
\[
\begin{array}{rclcrcl}
W_0 & \! = \! & \left(
\begin{array}{cccc}
1 & 0 & 0 & 0 \\
0 & 1 & 0 & 0 \\
0 & 0 & 1 & 0 \\
0 & 0 & 0 & 1 
\end{array}
\right)
& & W_1 & \! = \! & \left(
\begin{array}{cccc}
0 & 0 & 0 & 0 \\
1 & 0 & 0 & 0 \\
0 & 0 & 0 & 0 \\
0 & 0 & 1 & 0 
\end{array}
\right)
\\
& & & & & & 
\\
W_2 & \! = \! & \left(
\begin{array}{cccc}
0 & 0 & 0 & 0 \\
0 & -2 & 0 & 0 \\
1 & 0 & -2 & 0 \\
0 & 0 & 0 & 0 
\end{array}
\right)
& & W_3 & \! = \! & \left(
\begin{array}{cccc}
0 & 0 & 0 & 0 \\
0 & 0 & 0 & 0 \\
0 & 0 & 0 & 0 \\
1 & 0 & -2 & 0 
\end{array}
\right)
\, .
\end{array}
\]
The dual Hopf algebra $\A$ is spanned by $\{f^i\}=\{1,a,b,ab\}$,
with commutation relations $a^2=0$, $b^2=-2b$ and $ba=-ab-2a$.
The Hopf structure is
\bae
\Delta(a) &=& a\ot 1 + 1 \ot a + b \ot a
\ff
\Delta(b) &=& b \ot 1 + 1 \ot b + b \ot b
\, 
\label{Hopf2dqs}
\eae
and $\epsilon(a)=\epsilon(b)=0$, $S(a)=a(1+b)$, $S(b)=b$ (so that
$S^2(a)=-a$). 
\end{example}

\noindent Background material on Hopf algebras can be found
in~\cite{SweeHA} and~\cite{Majid1}. 
\section{Quantum Integrals}
\label{QInt}
\subsection{A trace formula}
\label{tf}
Having codified translations in the coproduct, we now
define a {\em right integral} in the Hopf algebra $\A$ as a map from 
functions to numbers, 
$\Rolo{\cdot}:$ $\A \ra \CC$, which  is invariant under right
translations
\be
\la a_{(1)} \Rar a_{(2)} = \la a \Rar \uA 
\label{rint}
\ee
for all $a$ in $\A$. 
We call $\la \cdot \Rar$ {\em trivial} if all
$\la f^{i} \Rar$ are zero. {\em Left integrals} are
similarly defined via
\be
a_{(1)} \la a_{(2)} \Lar = \uA \la a \Lar 
\, . 
\label{lint}
\ee
Radford and Larson~\cite{LarRad1} give the following  trace 
formula for the right integral
\be
\la a \trRar \equiv \ip{S^{2}(e_{i})}{f^{i}a}
\, , 
\label{RadL}
\ee
where $e_i \ot f^i$ is the canonical element, 
as they warn though, it does not always produce a non-trivial
result. 
\begin{example}{Integration on discrete groups}
\label{idg}
Let's try~(\ref{RadL}) in a classical setting.
$S^2 = \id$ in this case and $ \la a \trRar=
\ip{e_i}{f^ia}= \sum_if^i(e_i) a(e_i) =\sum_ia(e_i)$ which is the
standard formula for the integral on the discrete group $\H$, 
normalized so that $\la \uA \ar_{tr}^{\scr R}=|\H|$.
\end{example}

Notice that~(\ref{RadL}) defines a (possibly trivial) invariant
integral for {\em every} Hopf algebra, without requiring that it be of
the function type. Consequently, we can use it to evaluate the
integral of a group element or a difference operator. What is, 
classically,  the meaning of
such an integral? Just like the integral of a function is the sum
of its pairings with all the elements of a basis in the dual, the
integral of a group element is the sum of the values, on that
particular element,  of all the
functions in a basis in the dual,  and likewise for a
difference operator.   
\begin{example}{Integral of a group element}
\label{Ige}
We use again~(\ref{RadL}), with $S^2=\id$, to compute the integral
of a group element $e_i$. We have $\olo{e_i}=\ip{e_i
e_j}{f^j}={M_{ij}}^k \ip{e_k}{f^j} = {M_{ij}}^j = {\delta_i}^0$,
since from $e_i e_j = e_j$ it follows that $e_i = e_0 =1$. For the
difference operators $e_{ij}$ we get then $\olo{e_{ij}}=\delta_{i0}$. 
\end{example}
\begin{example}{Failure of the trace formula}
\label{ftform}
We compute the integral given by~(\ref{RadL}) for the Hopf
algebra of Ex.~\ref{dqs}.
Using the definition of $W^k$ given in the above example
we find from~(\ref{RadL})
\be
\la f^k \trRar  = \mbox{Tr}(S^2 W^k)
\, .
\label{RadLnum}
\ee
Inspection of the matrices given explicitly in~(\ref{MWdqs})
shows that $\la f^k \trRar=0$, $k=0,\dots,3$.   
\end{example}

\noindent Further material for this section can be found
in~\cite{LarSwee,Rad,VD1}.
\subsection{A non-trivial trace formula}
\label{nttf}
We want to analyze now under what conditions does~(\ref{RadL})
fail and try, if possible, to modify it so that it furnishes
always a non-trivial result. Our approach will be based on a
`vacuum expectation value' treatment of the integral~\cite{BZpc}. 
To arrive at
such a formulation, we first remark that the invariance
relation~(\ref{rint}), when paired with an arbitrary $x$ in $\H$
gives
\be
\la x \tr a \ar = \epsilon(x) \la a \ar 
\, , 
\label{rint2}
\ee
where $x \tr a \equiv a_{(1)} \ip{x}{a_{(2)}}$. For $x$ a group-like
element, $x \tr a$ is the function $a$ (right) translated by $x$.
When $x$ is a difference-like operator, $x \tr a$ is the
`derivative' of $a$ along $x$. In the first case, $\epsilon(x)=1$
and~(\ref{rint2}) 
states that the integral of the translated function is equal to 
that of the original one while in the second, $\epsilon(x)=0$ and
we get that the integral of a derivative is zero. 

Given that $x$
acts on products of functions via its coproduct, $x \tr (ab)=
(x_{(1)} \tr a)(x_{(2)} \tr b)$, one can form a new algebra $\A
\rtimes \H$, the {\em semidirect product} of $\A$ and $\H$, 
containing these as subalgebras and with cross relations 
\bae
xa &=& (x_{(1)} \tr a) x_{(2)}
\nonumber \\
 &=& a_{(1)} \inprod{x_{(1)}}{a_{(2)}} x_{(2)}
\, .
\label{AUcr}
\eae
The inverse relations are
\be
ax = x_{(2)} \inprod{x_{(1)}}{S^{-1}(a_{(2)})} a_{(1)} \, .
\label{AUcr1}
\ee
Consider now two formal symbols $\urv$ and $\rva$, the $\U$ and
$\A$-{\em right vacua} respectively~\cite{CZ}, which satisfy
\bae
x \urv \fe \epsilon(x) \urv \ff
a \rva \fe \epsilon(a) \rva .
\label{HAvac}
\eae
{\em Left vacua} $\ulv$, $\lva$ are defined analogously.
In terms of these, the inner product $\inprod{x}{a}$ can be given
as the `expectation value'
\bae
\lva xa \urv \fe \lva a_{(1)} \inprod{x_{(1)}}{a_{(2)}} x_{(2)}
\urv \ff
 \fe \inprod{x}{a}
\, ,
\nn
\eae
if we normalize the vacua so that $\langle \vaca | \vacu \rangle
=
\langle \vacu | \vaca \rangle = 1$. Similarly, the left action
of $\U$ on $\A$ can be written as
\be
x \tr a \urv = xa \urv \nn \, . 
\label{xacrU}
\ee
We can now write a symbolic expression for our `vacuum' integral
\be
\la a \ar_{v}^{\scr R} \sim \rva \ulv a \urv \lva \label{defofint} .
\ee
Invariance in the form~(\ref{rint2}) follows from~(\ref{xacrU})
and the left version of the first of~(\ref{HAvac}). 
This can be turned into something more than symbolic if we borrow
from~\cite{CSW} the result that the operators $\rva \ulv$, $\urv
\lva$ can be represented in $\A \rtimes \H$ as
\bae
\urv\lva &\sim& E \equiv  S^{-1}(f^{i}) e_{i} \, , 
\ff
\rva \ulv &\sim&  \bar{E}  \equiv S^{2}(e_{i}) f^{i}
\, .
\label{EbEdef}
\eae
Indeed, one easily finds that $Ea=\epsilon(a) E$, $xE=\epsilon(x)
E$ \etc, as well as that $E^2=E$, $\bar{E}^2= \bar{E}$. 
Then~(\ref{defofint}) can be given a precise meaning  by 
defining~\cite{CC}
\be
\la a \ar_{v}^{\scr R} \rva \lva \equiv \bar{E} a E
\, .
\label{intvdef}
\ee
This needs perhaps some explanation. The r.h.s. above is an 
element of $\A
\rtimes \H$ that realizes the operator $\rva \lva$. The latter is
unique up to scale, and hence, for different inputs $a$ in the
l.h.s., we get numerical multiples of the same operator in the
r.h.s. (after
using the commutation relations in $\A \rtimes \H$ to bring the
r.h.s. in some standard ordering). 
After choosing a 
normalization for $\rva \lva$, we may define $\la a \ar_{v}^{\scr
R}$ to be
the numerical constant that multiplies $\rva \lva$ in the l.h.s. above. 
Notice that when the left and right integrals coincide, 
$\bar{E} a E$ is a pure function and $\rva \lva$ is realized in
$\A$.
The connection of this definition with the trace formula is
revealed by computing
\bae
\bar{E} a E \fe S^{2}(e_{i}) f^{i} a E \ff
 \fe f^{i}_{(1)} a_{(1)} \ip{S^{2}(e_{i})}{f^{i}_{(2)} a_{(2)}} E
\ff
 \fe f^{n} \ip{e_{n}}{f^{i}_{(1)} a_{(1)}}
\ip{S^{2}(e_{i})}{f^{i}_{(2)} a_{(2)}} E \ff
 \fe f^{n} \ip{e_{n} S^{2}(e_{i})}{f^{i}a} E 
\, . \label{EbaraE}
\eae
The quantity
multiplying $E$ in the last line above is a modified trace which
is both right-invariant and non-trivial.
Indeed, invariance is easily verified along the lines of the
proof of the  original trace formula, Eq.~(\ref{RadL}). For the 
non-triviality, we
set $\Theta^{i}_{l} \equiv \ip{e_{i} S^{2}(e_{k})}{f^{k} f^{l}}$
and compute
\bae
S^{2}(e_{i}) f^{i} \fe f^{i}_{(1)}
\ip{S^{2}(e_{i_{(1)}})}{f^{i}_{(2)}} S^{2}(e_{i_{(2)}}) \ff
 \fe f^{i}_{(1)} f^{j}_{(1)} \ip{S^{2}(e_{i})}{f^{i}_{(2)}
f^{j}_{(2)}} S^{2}(e_{j}) \ff
 \fe S^{2}(f^{i}) f^{j} \ip{e_{i} S^{2}(e_{k})}{f^{k} f^{l}}
\ff
 & & \qquad \qquad S^{2}(e_{j}) S^{2}(e_{l}) \ff
 \fe \Theta^{l}_{i} S^{2}(f^{i}) f^{j} S^{2}(e_{j})
S^{2}(e_{l})
\label{nontriv1}
\eae
from which we may conclude that not all $\Theta^{l}_{i}$ are
zero, since $S^{2}(e_{i})f^{i} \neq 0$ (see~\cite{VD1} for an
alternative proof). Some Hopf algebra
trickery shows that $\ip{xS^2(e_i)}{f^ia}$ supplies actually a
left integral for $x$ (as well as a right one for $a$), so it is
proportional to their product. 
Then~(\ref{RadL}) returns the integral of $a$ multiplied by the
integral of the unit element in the dual and will hence fail
whenever the latter vanishes. Our modified trace above avoids
this by summing over all integrals in the dual and `tagging' each
by the (linearly independent) $f^n$. 
The appearance of $\rva \lva$ in~(\ref{intvdef}) might look a 
little strange now, but
it is actually a blessing in disguise. When later we look for an
integral in braided Hopf algebras, it will become apparent  
that a purely numerical integral cannot, in general,  transform
covariantly and the operator $\rva \lva$ that the braided analog of the
above construction produces exactly compensates for the fact that
we do not explicitly display the measure used in our integrations. 
\begin{example}{The modified trace}
\label{mtr}
Continuing Ex.~\ref{dqs}, we derive the commutation relations
of~(\ref{AUcr})
\bae
xa &=& 1+b +ax
\ff
xb &=& -(b+2)x
\ff
ya &=& ay
\ff
yb &=& 1+b- (b+2)y
\, .
\label{AUcrdqs}
\eae
We also compute the vacuum projectors
in~(\ref{EbEdef})
\bae
E &=& 1 -ax(1-2y) + by -abx(1-y)
\ff
\bar{E} &=& 1 -xa +yb-xyab
\, .
\label{EbEdqs}
\eae
One easily verifies that they satisfy indeed $Ea=\epsilon(a)E$
\etc .
A straightforward calculation now gives
\bae
\la \uA \ar_{v}^{\scr R}  \, \rva \lva &=& \bar{E} E = 0
\ff
\la a \ar_{v}^{\scr R} \, \rva \lva &=& \bar{E} a E = -abg
\ff
\la b \ar_{v}^{\scr R} \, \rva \lva &=& \bar{E} b E = 0
\ff
\la ab \ar_{v}^{\scr R} \, \rva \lva &=& \bar{E} ab E = abg
\, 
\label{EbEab}
\eae
where $g \equiv 1-2y$. The element $abg \in \A \rtimes \H$
realizes (up to scale) the operator $\rva \lva$. We fix  
the normalization so
that $\la a \ar_{v}^{\scr R} = -1$ and $\la ab \ar_{v}^{\scr R} = 1$, 
the other two
integrals being zero. We can also treat $E$ above as a spectator
to find $\bar{E} a E= -ab E$ and $\bar{E} ab E= ab E$, thus 
isolating the
right delta function $\Rda=ab$. A left invariant integral in $\A$
is given by $\la \cdot \ar_{v}^{\scr L} = \la S(\cdot) \ar_{v}^{\scr
R}$. We find
$\la ab \ar_{v}^{\scr L}=-1$, all other integrals  being zero. 

Noting that $x \triangleleft a \equiv \ip{x_{(1)}}{a} x_{(2)}$
satisfies $E x \triangleleft a = Exa$, we may compute a {\em
left}-invariant integral in $\H$ via 
\be
\la z \ar_{v}^{\scr L} \urv \ulv = Ez\bar{E} \qquad z \in \H
\, .
\label{HLint}
\ee
We find that the only non-zero integral is 
\bae
 \la z \ar_{v}^{\scr L} \urv \ulv &=& Exy\bar{E}
\ff
 & = & -(1+b)x(1-y)
\, ;
\label{HLxy}
\eae
notice that $\Lolo{z} \propto
\ip{z}{\Rda}$. The r.h.s. of~(\ref{HLxy}) realizes $\urv \ulv$
in $\A \rtimes \H$. A right integral in $\H$ is given, as
before, via the antipode.
\end{example}
\section{Hopf Algebras + Statistics\\= Braided Hopf Algebras} 
\label{HASBHA}
\subsection{The universal $R$-matrix}
\label{uRm}
It is a rule of thumb in algebra that 
the interesting way to generalize a symmetry or constraint is by
relaxing it `up to similarity'. 
Deforming the
coproduct of our `points' into a non-cocommu\-ta\-ti\-ve one gives rise
to a rich algebraic structure when the two coproducts
$\Delta$ and $\Delta' \equiv \tau \circ \Delta$ (with $\tau$ the 
permutation map) are related by conjugation 
\be
\Delta ' (x) = {\cal R} \Delta (x) {\cal R}^{-1}  \qquad \qquad
\label{RcopRi}
\ee
for all $x$ in $\U$. Here $\R \in \H \ot \H$ is the {\em
universal $R$-matrix} which satisfies additionally
\bae
(\Delta \otimes \id ){\cal R} \fe {\cal R}_{13} {\cal R}_{23}
\label{Rprop1} \\
( \id \otimes \Delta) {\cal R} \fe {\cal R}_{13} {\cal R}_{12}
\, .
\label{Rprop2}
\eae
A Hopf algebra $\H$ for which an $\R$ exists is called {\em
quasitriangular}. Consider the collection of (left) representation
spaces  of such an $\H$ (left $\H$-{\em mo\-du\-les}). These are
vector spaces (possibly with additional structure, like product,
coproduct \etc) on which $\H$ acts from the left.
This collection is equipped with a {\em tensor product} operation that 
combines two $\H$-modules $V$, $W$ to form a new one, their {\em
tensor product} $V \ot W$, on which $\H$ acts via its coproduct
\be
h \tr (v \ot w) = (h_{(1)} \tr v) \ot (h_{(2)} \tr w)
\, .
\label{HactVW}
\ee
Another way to obtain new modules from old ones, at least in the
classical case, is via the transposition $\tau:$ $V \ot W \mapsto
W \ot V$. This operation commutes, in the classical case, with
the action of $\H$ since $\Delta' = \Delta$. To obtain an
analogous operation in the Hopf algebra case, we need to employ a
{\em braided transposition} $\Psi$ which consists in first acting
with $\R$ on the tensor product of the modules to be transposed
and then effecting $\tau$
\bae
\Psi(v \ot w) &\equiv& \tau(\R^{(1)} \tr v \ot \R^{(2)} \tr w)
\ff
 & = & \R^{(2)} \tr w \ot \R^{(1)} \tr v 
\, ,
\label{Psidef}
\eae
where we have written $\R \equiv \R^{(1)} \ot \R^{(2)}$
(summation implied). Things get interesting when $\R' \equiv
\tau(\R) = \R^{(2)} \ot \R^{(1)} \neq \R^{-1}$, which is often
the case. This results in $\Psi^2 \neq \id$ (by $\Psi^2$ we mean
$\Psi_{V \ot W} \circ \Psi_{W \ot V}$). It is as though the
objects we transpose were hanging from the ceiling by threads and
successive transpositions  were recorded in the entanglement of the 
threads. Taking this picture seriously, we will adopt a
diagrammatic notation in which $\Psi$, $\Psi^{-1}$ look like
\[
\raisebox{0\totalheight}[\height]{%
\epsfig{file=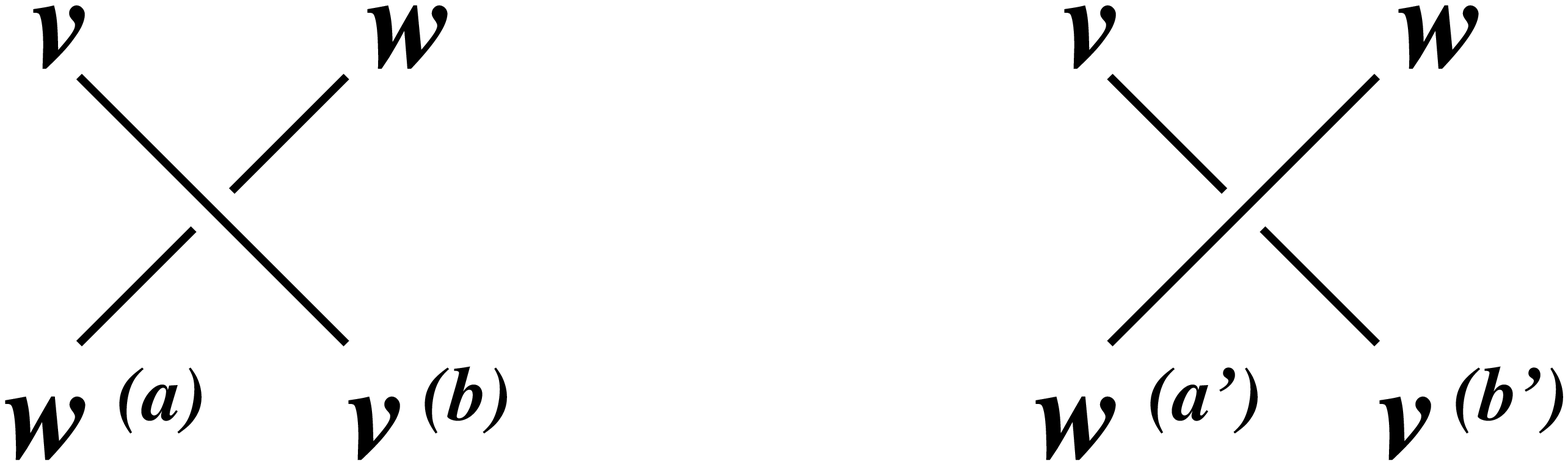,width=30ex} 
}
\]
where $\Psi(v \ot w) \equiv w^{(a)} \ot v^{(b)}$ and 
$\Psi^{-1}(v \ot w) \equiv w^{(a')} \ot v^{(b')}$. In
other words, we imagine the algebra elements flowing from top to
bottom along the braids, with over- and under-crossings
representing the effect of $\Psi$ and $\Psi^{-1}$ respectively.
We see that the existence of an $\R$ in $\H$ endows all algebraic 
structures
covariant under the action of $\H$ with a natural {\em
statistics}. The usual bosonic and fermionic rules for
transposition can be put in this language but, in general, $\Psi$
can be much more drastic, as the action of $\R$ that precedes
that of $\tau$ if often far from trivial. 
\subsection{Braided Hopf algebras}
\label{BHa}
One might think now of developing an algebra, involving products,
coproducts \etc, in which all typographical transpositions are
effected by $\Psi$, $\Psi^{-1}$, rather than $\tau$. {\em Braided 
Hopf algebras} are a transcription of ordinary Hopf algebras in
this braided setting~\cite{Majid4}. Thus, one has a braided 
antipode, braided
coproduct \etc, which satisfy brai\-ded versions of the standard
Hopf algebra axioms. Denoting \eg \ the product and coproduct by
the first two of the vertices below
\[
\raisebox{0\totalheight}[1.1\height][.1\height]{%
\epsfig{file=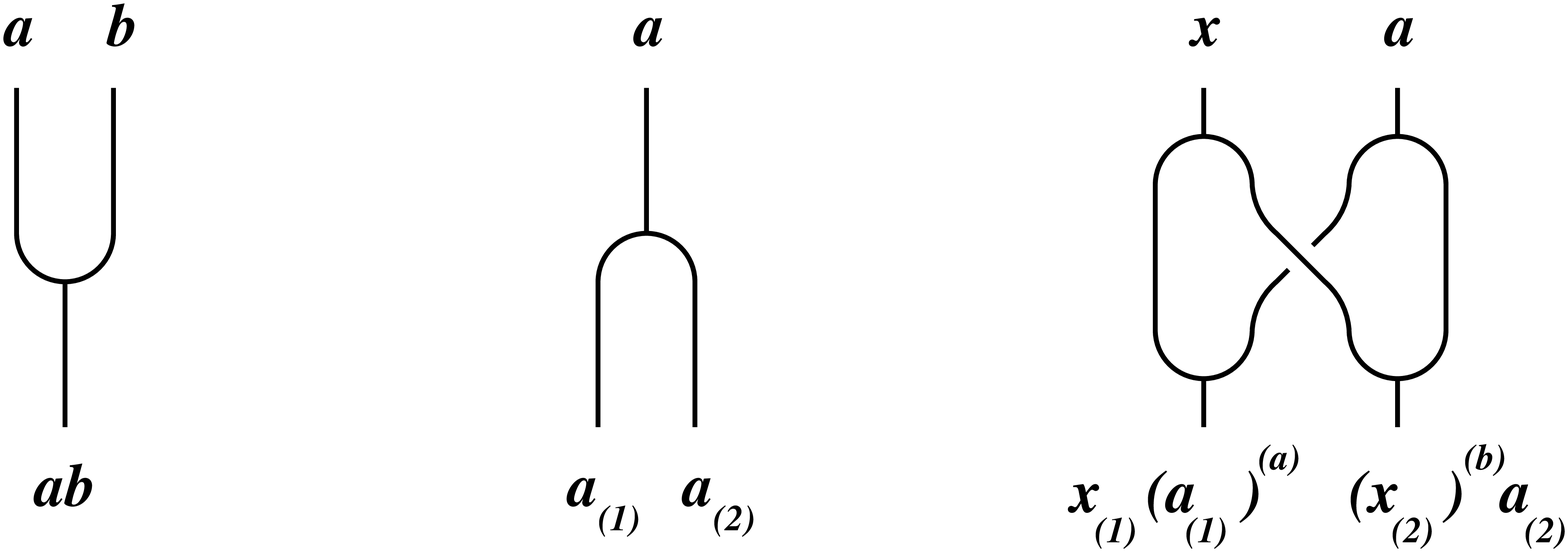,width=45ex}}
\]
one gets the braided version of the multiplicativity
of the coproduct expressed by the third diagram above. Similarly,
we denote $\epsilon$, $S$, $S^{2}$, $S^{-1}$, $S^{-2}$
respectively by
\[
\raisebox{0\totalheight}[1.1\height][.1\height]{%
\epsfig{file=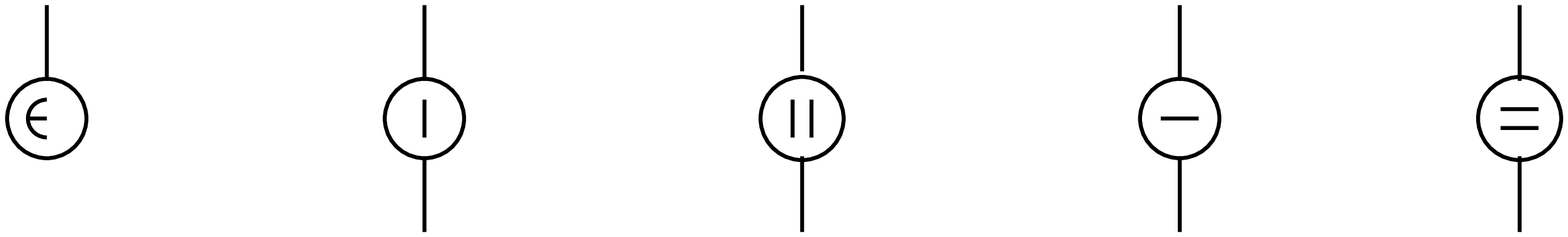,width=45ex}}
\]
and express the braided analogues of \eg 
\bae
a_{(1)} \epsilon(a_{(2)}) &=& a
\ff
S(a_{(1)}) a_{(2)} &=& \epsilon(a)
\ff
S(ab) &=& S(b)S(a) 
\nn
\eae
by the diagrams
\[
\raisebox{0\totalheight}[\height][0\height]{%
\epsfig{file=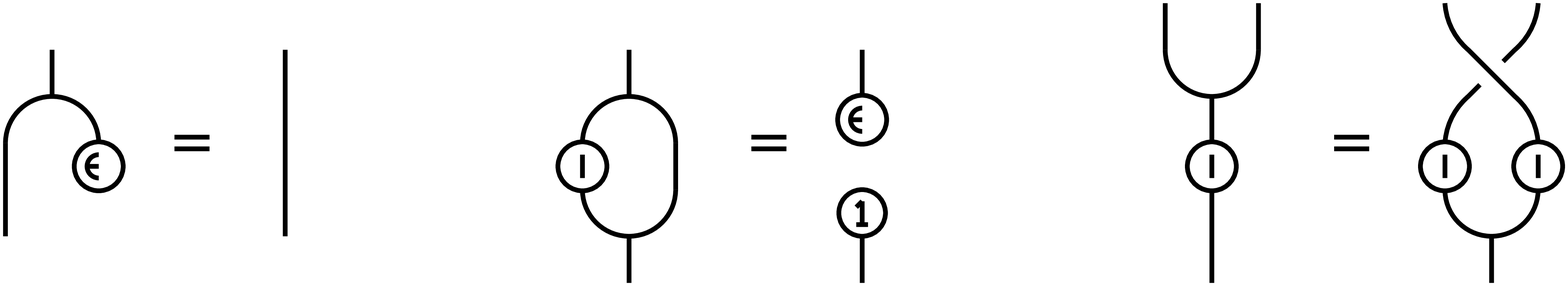,width=45ex}}
\]
respectively (the output braid of the counit is supressed since,
being a number, it braids trivially). Covariance is codified in the 
requirement
that one should be able to move crossings past all vertices and
boxes, \eg \ the relations 
\[
\raisebox{0\totalheight}{%
\epsfig{file=funct.eps,width=48ex}}
\]
should hold.  The inner product and the canonical
element look like  
\[
\raisebox{0\totalheight}[\height][-.1\height]{%
\epsfig{file=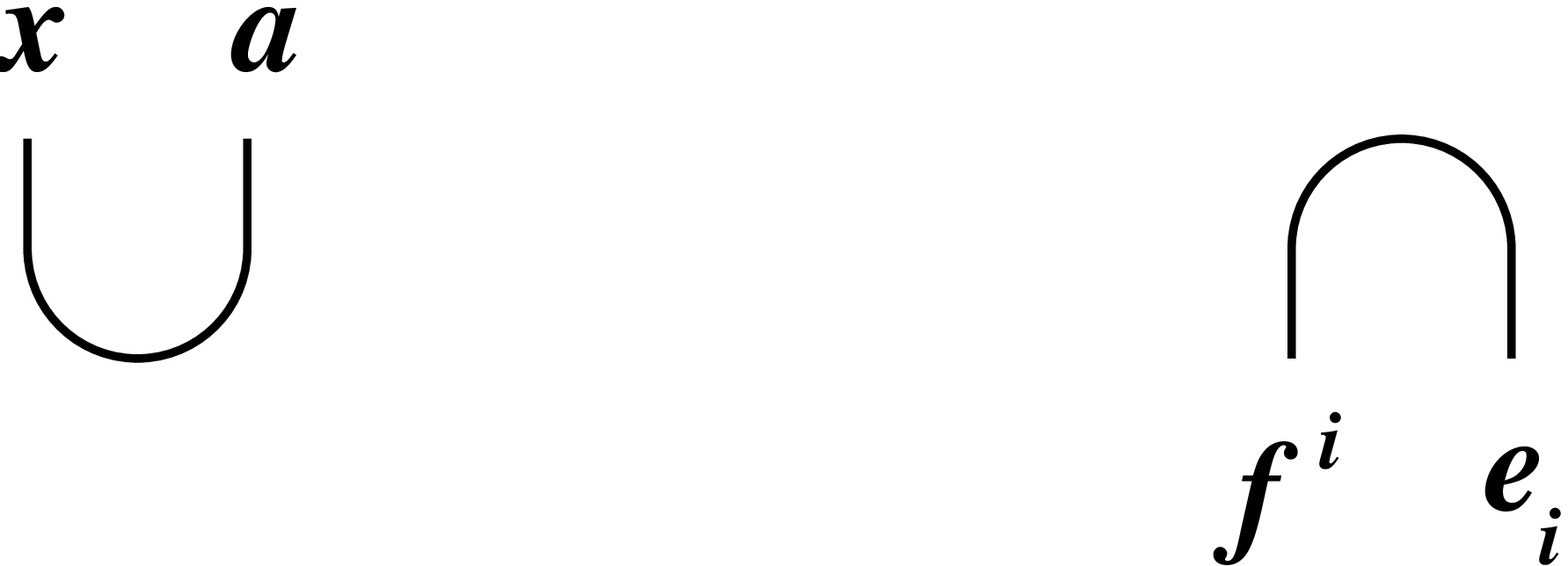,width=25ex}}
\nn
\]
For the product-coproduct duality we adopt a convention which is
slightly different from the one we used before, as shown below
\[
\raisebox{0\totalheight}[1.2\height][1.2\depth]{%
\epsfig{file=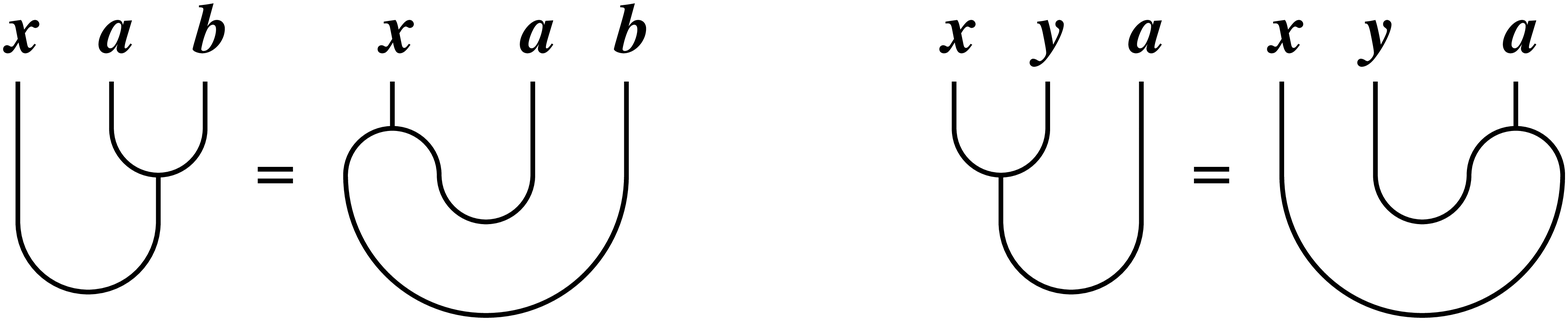,width=47ex}}
\]
As a result, to get the unbraided expression that corresponds to
any of the diagrammatic identities that appear in the following,
one should translate the diagrams, ignoring the braiding
information, into the language of Sect.~\ref{QInt} and then set
$\Delta \rightarrow\Delta'$, $S \rightarrow S^{-1}$. Notice that
all diagrams reveal new (dual) information when viewed
upside down. 
The commutation relations in the semidirect product (\ie \ the
braided analogue of (\ref{AUcr})) are
\[
\raisebox{0\totalheight}[1.2\height][1.2\depth]{%
\epsfig{file=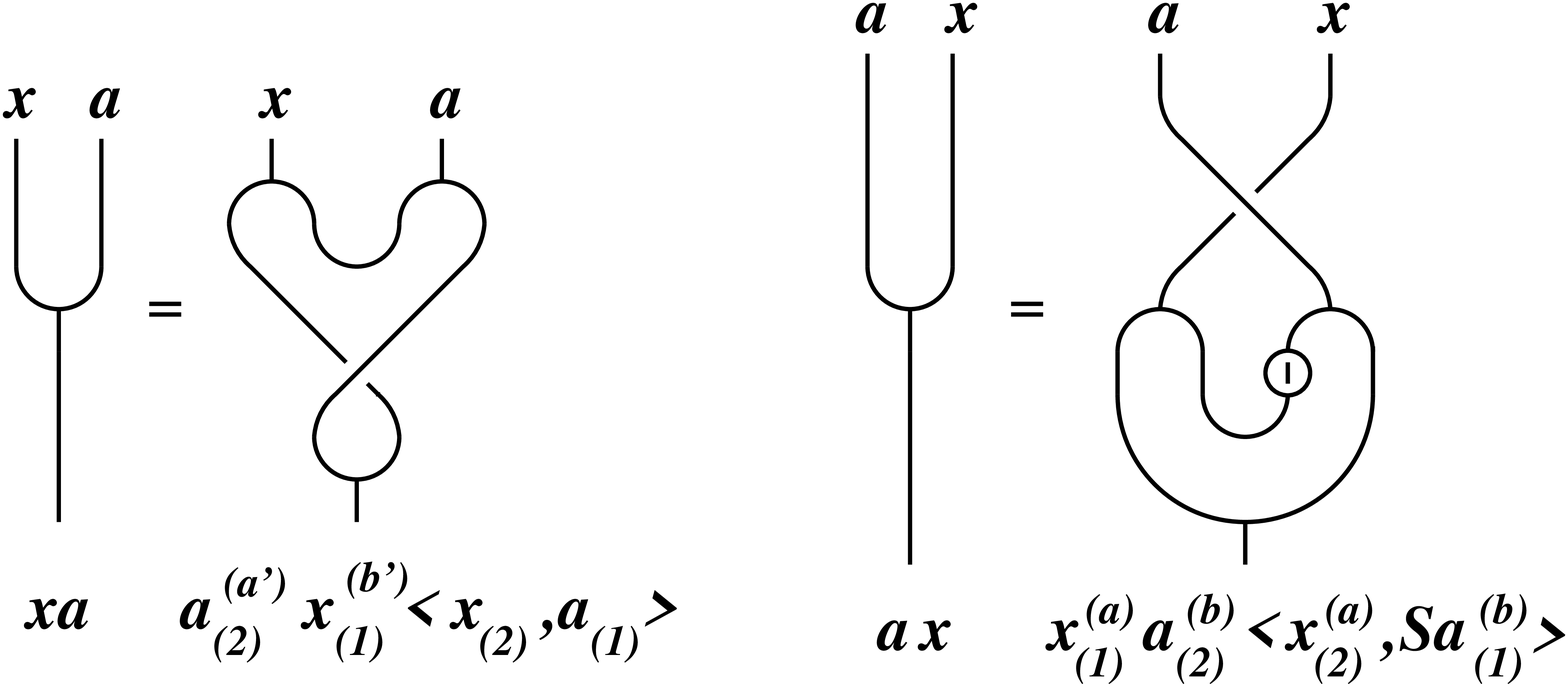,width=47ex}}
\]
A detailed exposition of the basics (and more) of
braided Hopf algebras can be found in~\cite{Majid4} and
references therein. 
\section{Braided Integrals} 
\label{BInt}
\subsection{Problems with braiding}
\label{Btf}
In the case of braided Hopf algebras, the integral presents an
additional complication. Defining it, along the lines of~(\ref{rint}), 
as a number, implies that it braids trivially, regardless of the
transformation properties of the integrand. This can easily be
seen to lead to problems when combined with the translational
invariance property, as the following toy example shows.
\begin{example}{The fermionic line}
\label{fermi1}
Consider the algebra $\A$ of functions on the
classical fermionic line. It is generated by $1$ and a
fermionic variable $\xi$, with $\xi^{2} =0$, and admits the
braided coproduct $\Delta(\xi) = \xi \ot 1 + 1 \ot \xi$, with
$\Psi$ the standard fermionic braiding, $\Psi(\xi \ot \xi)= -\xi
\ot \xi$ ($\Psi^2=\id)$ in this case). $\epsilon(\xi)=0$ and
$S(\xi)= -\xi$. Representing the integral with a rhombus, we want
it to satisfy
\be
\raisebox{-.5\totalheight}[.6\totalheight][.6\totalheight]{%
\epsfig{file=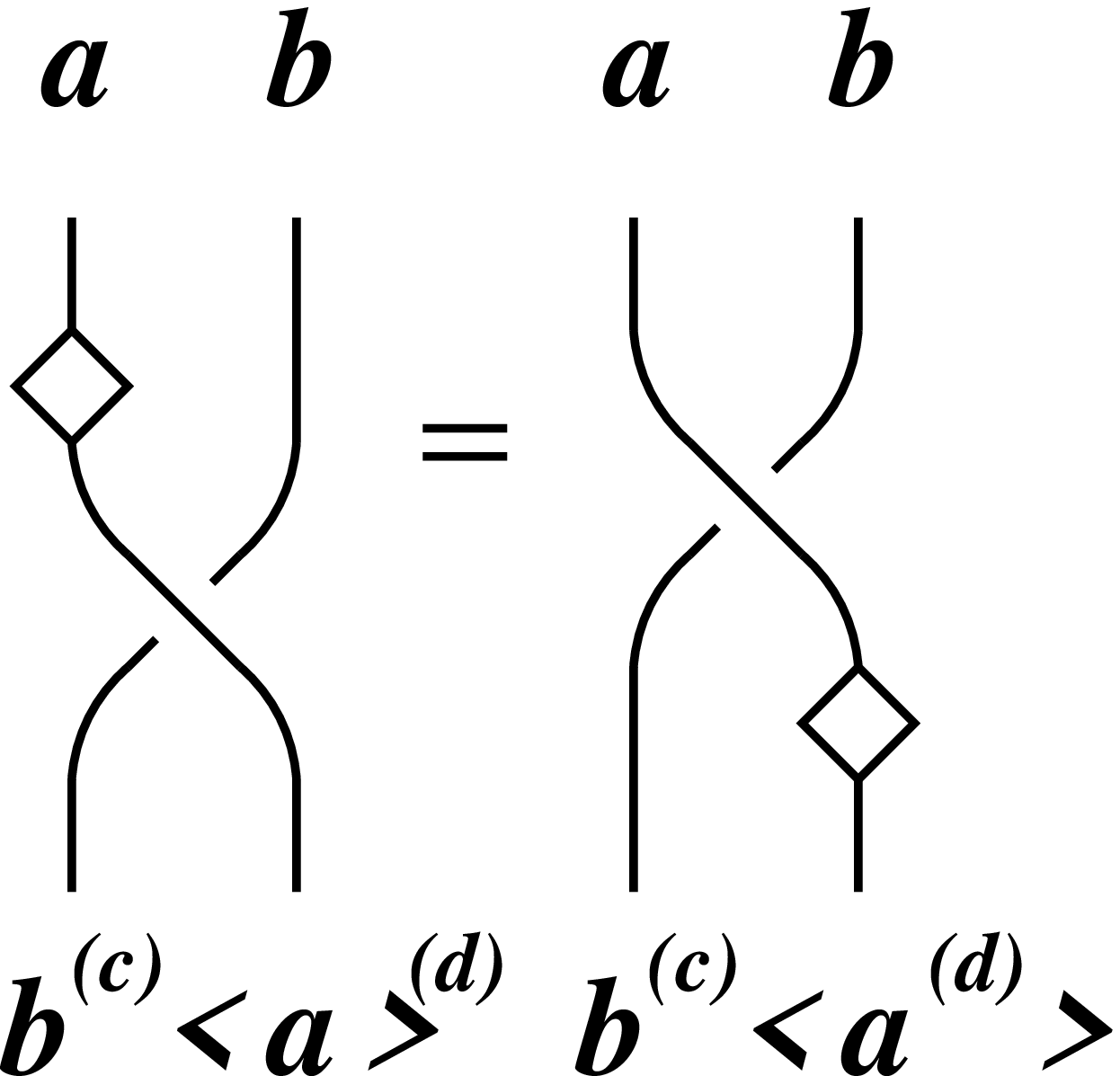,width=18ex}}
\label{functint1.fig}
\ee
Taking both inputs to be $\xi$, and using the Berezin result
$\olo{\xi}=1$, we get for the l.h.s.
\bae
\Psi(\olo{\xi} \ot  \xi) &=& \Psi(1 \ot \xi)
\ff
 &=& \xi \ot 1
\nn
\eae
while the r.h.s. gives
\bae
(\id \ot \olo{\cdot}) \circ \Psi(\xi \ot \xi) &=& 
- (\id \ot \olo{\cdot}) (\xi \ot \xi)
\ff
 &=&- \xi \ot 1
\, .
\nn
\eae
The problem originates in the absense of any explicit mention, in
our algebraic treatment,  to the `measure' $d\xi$.
\end{example} 

\noindent More material on braided integrals is 
in~\cite{BKLT,Lyu1,Lyu2}.
\subsection{The braided vacuum projectors}
\noindent 
As we mentioned at the end of Sec.~\ref{QInt}, the `vacuum
expectation value' approach suggests a solution~\cite{CC}. 
Our starting point is the braided analogue of the vacuum
projectors of~(\ref{EbEdef}). Denoting them by $\E$, $\bE$ we
find
\be
\E \, = \, 
\raisebox{-.5\totalheight}{%
\epsfig{file=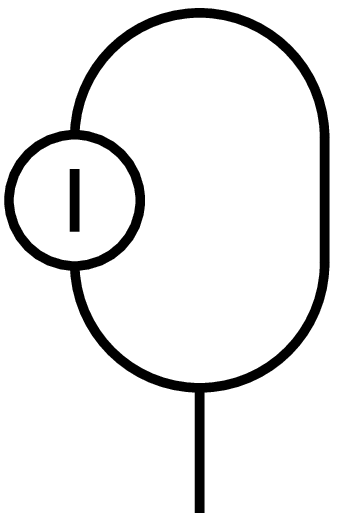,width=5ex}}
\qquad \qquad 
\bE \, = \, 
\raisebox{-.5\totalheight}{%
 \epsfig{file=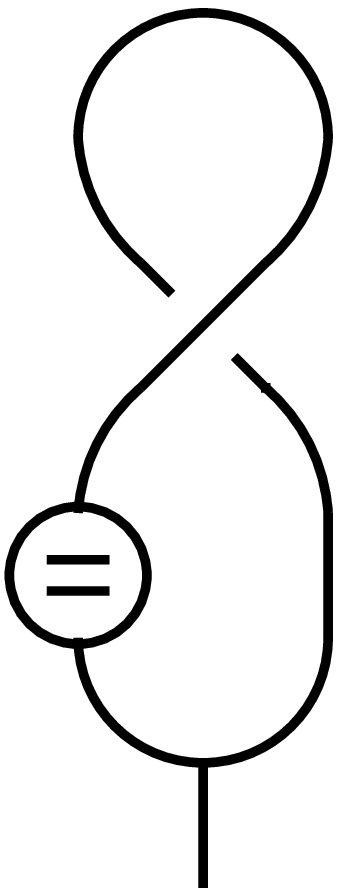,width=5ex}}
\label{bEEb}
\ee
We give, in Fig. \ref{Eproof}, the proof that $\E a=\E\epsilon(a)$, 
for all $a$ in $\A$---the rest of the required relations are
proved similarly. 
\FIGURE{%
\label{Eproof}
\raisebox{0\totalheight}[1.2\height][1.2\depth]{%
\epsfig{file=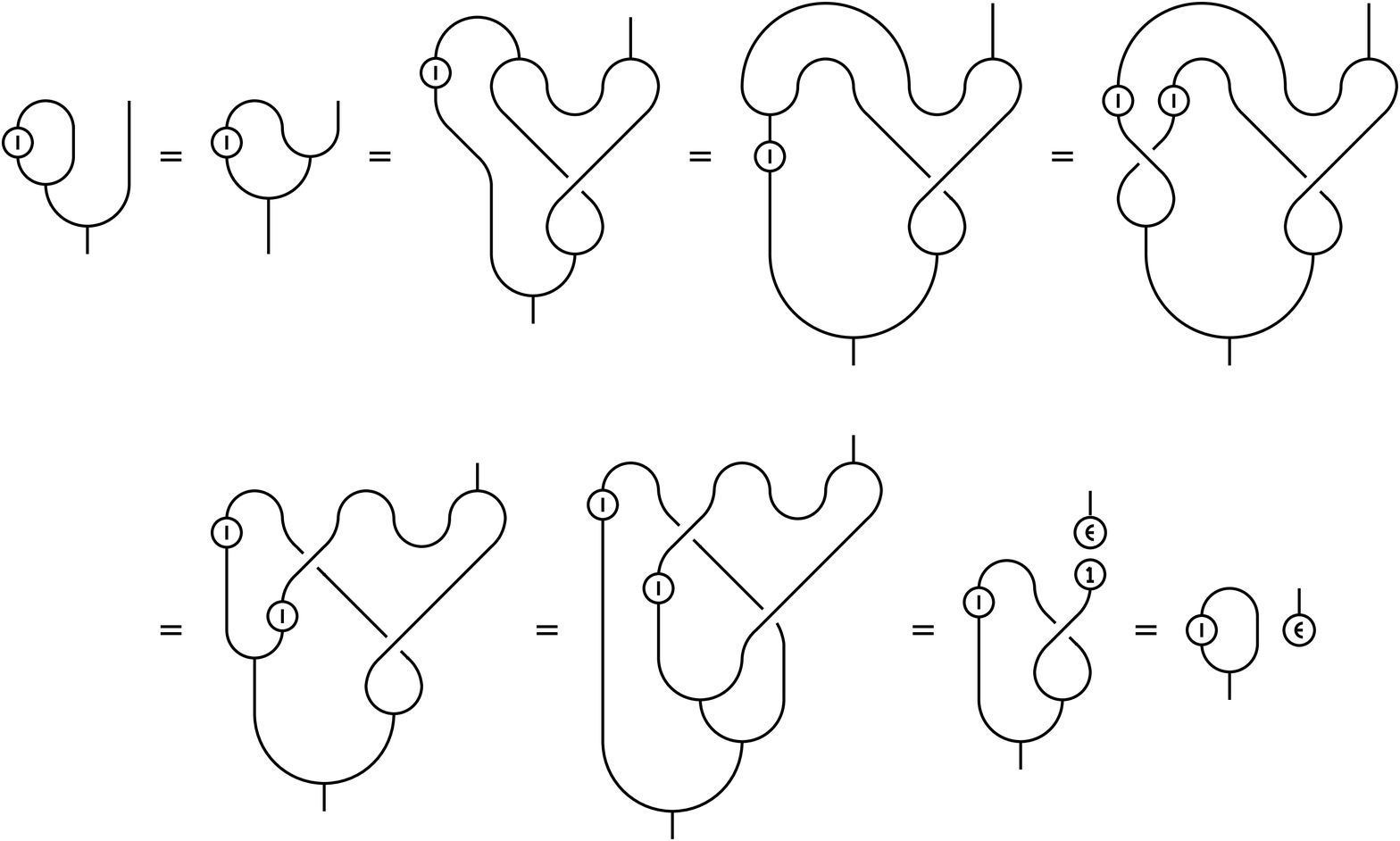,width=70ex}}
\caption{Proof of $\E a=\E\epsilon(a)$, $a\in \A$}
}
Forming $\bE a \E$ and 
treating $\E$ as a spectator, we get a glimpse of the inner workings
of the rhombus
\be
\raisebox{-.5\totalheight}[.6\totalheight][.6\totalheight]{%
\epsfig{file=rhoiw.eps,width=25ex}}
\label{rhoiw.fig}
\ee
The output in the above diagram is a multiple of the delta function, 
the
coefficient being the numerical integral of the input. Again, one
can show that this provides a non-trivial integral for all
finite dimensional braided Hpf algebras, transcribing 
either~(\ref{nontriv1})  or the more
elegant proof in~\cite{VD1}. Opting for the latter we find
\be
\raisebox{0\totalheight}[1.2\height][1.2\depth]{%
\epsfig{file=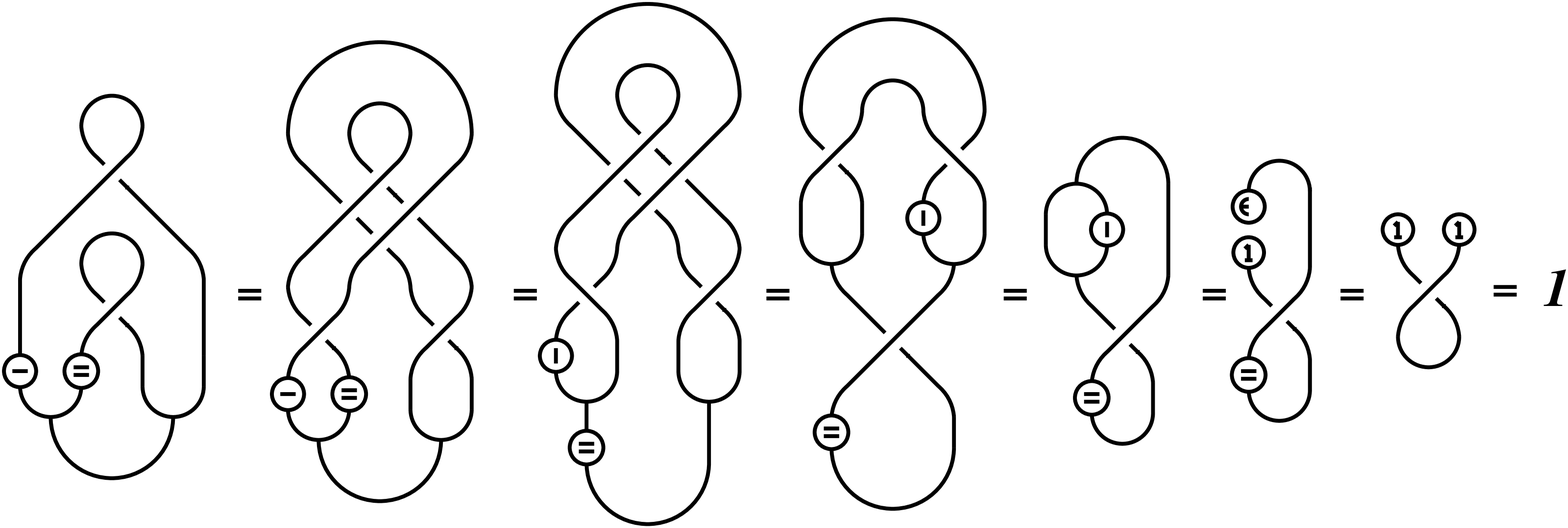,width=49ex}}
\label{nontriv.fig}
\ee
from which non-triviality follows. The analogue of~(\ref{RadL})
is easily deduced from~(\ref{rhoiw.fig})
\be
\la a \trRar \, = \, 
\raisebox{-.5\totalheight}[.5\totalheight][.5\totalheight]{%
\epsfig{file=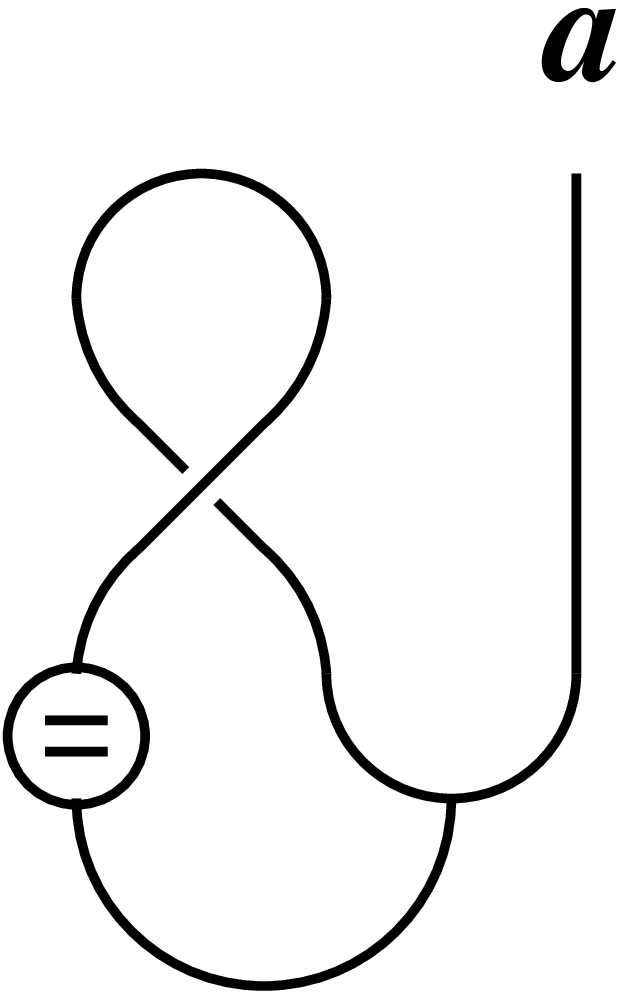,width=9ex}}
\label{bRL.fig}
\ee 
and a direct proof of its invariance is shown in
Fig. \ref{invar2.fig}. 
\FIGURE{\centerline{%
\label{invar2.fig}
\raisebox{0\totalheight}[1.2\height][1.2\depth]{%
\epsfig{file=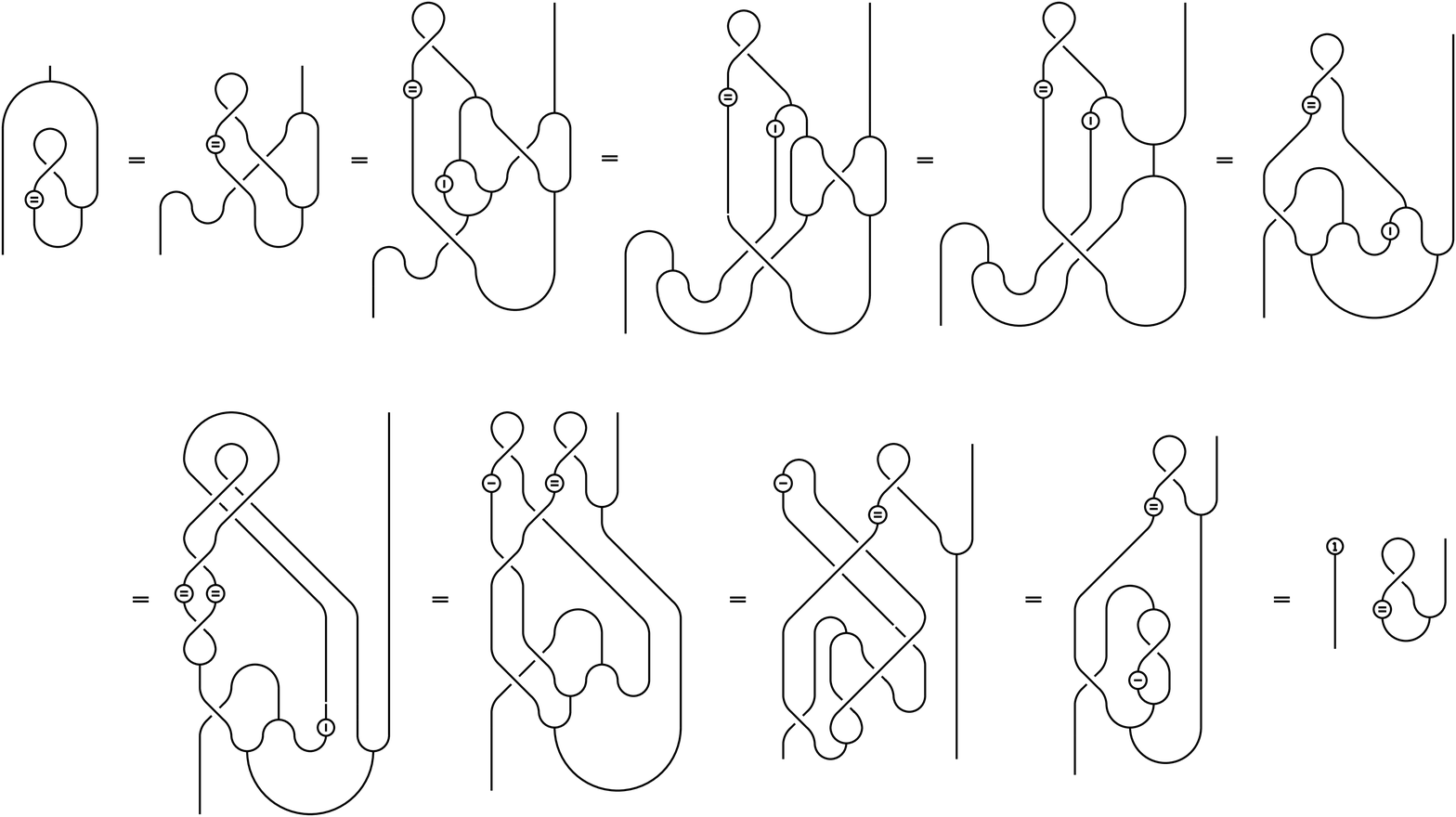,width=90ex}}}%
\caption{Proof of the invariance of the integral}
}
\begin{example}{The Berezin integral as a sum}
Continuing our toy  example~\ref{fermi1}, we introduce a fermionic
derivative $\sigma$, with $\sigma^2=0$ and $\sigma \xi =1 -\xi
\sigma$ and the standard Hopf structure and braiding. Then, 
$\{e_i\}=\{1,\sigma\}$ and $\{f^i\}=\{1, \xi\}$ which gives $\E=
1-\xi \sigma =\sigma \xi$ and $\bE=1-\sigma\xi = \xi \sigma$. For
the integrals we compute $\bE \E=0$ and $\bE \xi \E=\xi$. Left
and right integrals coincide in this case so $\rva \lva$ is
realized in $\A$ by $\xi$ and we recover the Berezin result.
Notice that~(\ref{rhoiw.fig}) gives the Berezin integral as a sum
over `points'. The integral of the unit function receives two
equal (unit)
contributions from the two `points' in the space, but the
undercrossing in $\bE$ flips the sign of one of them  so that
they cancel. 
\end{example} 
\begin{example}{The quantum fermionic plane}
This was introduced in~\cite{WZ,BZ2}---we follow the conventions
in \cite{CC}. 
$\A$ is generated by the fermionic coordinate functions $\xi_i$,
$i=1,\dots,N$. They satisfy
\be
\xi_{2} \xi_{1} = -q \hR_{12} \xi_{2} \xi_{1}    \label{xxcrs}
\ee
and are dual to the derivatives $\sigma_i$, $i=1, \dots ,N$ that
generate $\H$ with relations
\be
\sigma_{1} \sigma_{2} = -q \sigma_{1} \sigma_{2} \hR_{12} \, .
 \label{ddcrs}
\ee
In $\A \rtimes \H$ we have
\be
\sigma_i \xi_{j} = \delta_{ij} - q \hR^{-1}_{mj,ni} \xi_{n}
\sigma_{m} \, . 
\label{dxcrs}
\ee
The braided coproduct is $\Delta(\xi_i)=\xi_i \ot 1 + 1 \ot
\xi_i$ and similarly for the $\sigma$'s. The braiding is given by
\bae
\Psi(\xi_2 \ot \xi_1 ) &=& -q \hR_{12} \xi_2 \ot \xi_1
\ff
\Psi(\sigma_1 \ot \sigma_2 ) &=& -q  \sigma_1 \ot \sigma_2
\hR_{12}^{-1}
\ff
\Psi(\xi_i \ot \sigma_j) &=& -q^{-1} D_{la} \hR_{ia,bk}
D^{-1}_{bj} \sigma_l \ot \xi_k
\ff
\Psi(\sigma_i \ot \xi_j) &=& -q^{-1} \hR_{kj,li} \xi_l \ot
\sigma_k
\, .
\nn
\eae
For the canonical element we find
\[
f^{i} \ot e_{i} = e_{q^{-1}}(\xi_{i} \ot \sigma_{i})
\, ,
\]
(compare with the bosonic quantum plane result in~\cite{CZ}) where
\[
\begin{array}{rclcrcl}
e_{q}(x) &=& \sum_{k=0}^{\infty}\frac{1}{[k]_{q}!} x^{k} \, ,
&\quad & 
[k]_{q} &=& \frac{1-q^{2k}}{1-q^{2}}
\, ,
\\
 & & & & & & \\
\mbox{} [k]_{q}! & = & [1]_{q} [2]_{q} \ldots [k]_{q} \, , &
\quad & 
[0]_{q} &\equiv &1 
\, .
\end{array}
\]
The vacuum projectors are
\bae
\E \fe \sum_{k=0}^{N} \frac{(-1)^{k}}{[k]_{q^{-1}}!} \xs \sx 
\ff
\bE \fe \sum_{k=0}^{N} \frac{(-1)^{k} q^{k}}{[k]_{q}!}
D_{i_{1}j_{1}}
\ldots D_{i_{k}j_{k}} \sx \xi_{j_{1\dots k}} \, ,
\nn
\eae
where $\xs \equiv \xi_{i_1} \dots \xi_{i_k}$ and similarly for
$\sigma$.   
The integral of a monomial $\xi_{i_{1\dots r}}$, $r < N$, is given by
\[
\bE \xi_{i_{1\dots r}}  \E =
\left(\, \sum_{k=0}^{A}\frac{(-1)^{k} q^{k(k-2A+1)} [A]_{q}!}{[k]_{q}!
[A-k]_{q}!}\right) \xi_{i_{1\dots r}} \E
\, ,
\]
with $A \equiv N-r$. Using standard $q$-machinery, the sum in 
parentheses  can be shown
to vanish for $0\leq r < N$ while for $r=N$ the integrand
is (proportional to) the (left- and right-) delta function. Then 
both $\bE$ and $\E$ reduce to $1_{\A \rtimes \H}$ and the
numerical 
integral is (proportional) to 1. We conclude that the integral is
essentially independent of $q$. Notice that, in the limit $q \ra
-1$, the algebra~(\ref{xxcrs}) {\em does not} become a bosonic one,
\eg~(\ref{xxcrs}) implies $\xi_i^2=0$ which persists for all
values of $q$. 
\end{example}
\acknowledgments
Warm thanks are due to the organizers of the Summer Institute in
Corfu for their efforts, also for financial support. The event,
immersed in extraordinary physical beauty, blended 
the academic with the less so in artful
proportions. I thank in particular George Zoupanos for his
interest and support and Nicolaos Tracas for his efficient help. 

The research of the author is supported by the Spanish Ministry
of Education and Culture. 

\end{document}